\documentclass{article} % onecolumn (second format)
\usepackage{amssymb}
\usepackage{amsmath}
\usepackage{latexsym}
\usepackage{verbatim}
\usepackage[english]{babel}
\usepackage{graphicx}
\usepackage{esvect}
\usepackage{cite}
\usepackage{amscd}
\usepackage{epsfig}
\usepackage{pstricks,pst-node}
\usepackage[margin=1in]{geometry}

\newlength{\figurewidth}
\newlength{\enviropost}
\newcommand{\be}{\begin{equation}}
\newcommand{\ee}{\end{equation}}
\newcommand{\ble}[1]{\begin{equation} \label{#1}}
\newcommand{\bae}{\begin{eqnarray}}
\newcommand{\eae}{\end{eqnarray}}
\newcommand{\fle}[2]% 
{\vspace{1.5ex}
\be
\label{#1}
\mbox{%
\setlength{\fboxsep}{3ex}%
\framebox{$\dss #2 $}}
\ee} 
\newcommand{\flec}[2]%
{\vspace{1.5ex}
\be
\label{#1}
\mbox{%
\setlength{\fboxsep}{3ex}%
\framebox{$\dss #2 $}}
\, \, \, ,
\ee} 
\newcommand{\flep}[2]%
{\vspace{1.5ex}
\be
\label{#1}
\mbox{%
\setlength{\fboxsep}{3ex}%
\framebox{$\dss #2 $}}
\, \, \, .
\ee} 
\newcommand{\dss}{\displaystyle}

\newcommand{\calO}{\mathcal{O}}

\begin{document}

\title{Extracting Geometry from Quantum Spacetime:\\
Obstacles down the road}

\author{Yuri Bonder, Chryssomalis Chryssomalakos, and Daniel Sudarsky\\ \\
Instituto de Ciencias Nucleares, Universidad Nacional Aut\'onoma de M\'exico\\
Apartado Postal 70-543, Ciudad de M\'exico 04510, M\'exico\\ \\
bonder; chryss; sudarsky@nucleares.unam.mx }

\date{\today}

\maketitle
 
\begin{abstract}
Any acceptable quantum gravity theory must allow us to recover the classical spacetime in the appropriate limit. Moreover, the spacetime geometrical notions should be intrinsically tied to the behavior of the matter that probes them. We consider some difficulties that would be confronted in attempting such an enterprise. The problems we uncover seem to go beyond the technical level to the point of questioning the overall feasibility of the project. The main issue is related to the fact that, in the quantum theory, it is impossible to assign a trajectory to a physical object, and, on the other hand, according to the basic tenets of the geometrization of gravity, it is precisely the trajectories of free localized objects that define the spacetime geometry. The insights gained in this analysis should be relevant to those interested in the quest for a quantum theory of gravity and might help refocus some of its goals.
\end{abstract}

\section{Introduction}\label{Intro}
 
The recovery of classical spacetime from theories involving pregeometrical concepts, like Loop Quantum Gravity \cite{LQG}, Spin Networks \cite{SpinNetwork}, Causal Sets \cite{CaSets}, Causal Dynamical Triangulations \cite{CDT}, etc., is a highly nontrivial task. Here, we consider the problem of the recovery, at least at an effective level, of the standard geometrical notions of General Relativity (GR) from such quantum geometrical structures. Naturally, the spacetime metric that would emerge from a generic fundamental spacetime theory, 
should realize the equivalence principle, as that is the essence of GR.
 Namely, it should describe the motion of free physical objects without postulating it in an \textit{ad hoc} manner, that is, one should recover the world lines of the effective point-like objects, which must coincide with the geodesics of the metric. Otherwise we could not argue justifiably that we have recovered GR from a more fundamental theory.
 
Even if one learns how to build the space of physical states of the fundamental spacetime theory, and how to construct the operators associated with geometrical quantities, like light cones, four volumes, spatial volumes, areas, or lengths, it is essential that the spacetime metric derived form such notions plays the role that it has in the classical theory. In other words, the emerging spacetime geometry should be at the basis of the behavior of the physical objects, and it is precisely this feature, presumably a direct consequence of the fundamental theory, what should give empirical meaning to the notion of a spacetime metric. In a detailed analysis by Reichenbach \cite{Reichenbach}, it is shown that without such a clear connection between spacetime geometry and the matter one uses to probe it, the statements made about the former are vacuous. This, in turn, points towards an essentially relational approach \cite{Relational}. Something similar ought to hold, at a more fundamental level of description, if the standard classical characterization of spacetime is to emerge naturally from it. Thus, we take the view that the description of the matter fields, and their role in providing a characterization of the spacetime geometry, should be an essential aspect of such a theory.
 
The guiding principle of the analysis that we undertake is to associate suitable world lines to physical objects, which are taken to represent the geodesics of an emergent metric, assuming this is at all possible. Once this is accomplished, one could attempt to read off the effective metric from the collection of such world lines. We focus on the difficulties faced in such an enterprise which are ultimately tied to the fact that, in the underlying fundamental theory, one cannot expect to have a notion of point particles. In fact, the matter excitations are expected to have extended support since, when going from the fundamental theory towards an effective description, one would need to pass through a phase where matter is represented by quantum fields. The problem is then tied to the need to associate world lines to such extended excitations. We assume that such world lines, in the intermediate stages connecting the fundamental and effective descriptions, correspond to something like centers of mass (CoM) associated with relatively localized distributions of matter (i.e., regions with nonvanishing energy-momentum tensor).

The program thus involves associating to any extended matter distribution of a certain scale, a CoM world line, and using those world lines to extract the effective metric at that scale. Here we note that, as it is well known \cite{Papapetrou,Beiglbock}, even in GR, in general, the motion of extended objects does not correspond to geodesics of the underlying spacetime. This fact, in turn, is what suggests the idea that there might be a new effective metric for which such world lines were geodesics. However, as we shall see, trying to recover an effective spacetime metric from these CoM world lines is, at best, problematic.

It is worth mentioning that the question we explore is related to the problem of spacetime averaging. The issue of characterizing the spacetime average of relevant quantities in GR is an open problem (for a review see \cite{Hoogen:2009}). This problem was initially recognized in reference \cite{Shirokov:1963a} (an English version can be found in \cite{Shirokov:1963b}) and further attention was generated with the appearance of \cite{Ellis:1984}. Although there are proposals that work for particular cases, for example on scalars \cite{Raychaudhuri:1998}, or when the average is done on spatial hypersurfaces \cite{Buchert:2000}, and even some covariant versions for tensors \cite{Isaacson1968a,Isaacson1968b}, the issue is still unsettled. One of the open issues is the fate of the Einstein equations under such averaging procedure, since, in general, the Einstein tensor associated with the average metric is not equal to the average Einstein tensor, and thus, the connection of such objects with the average energy-momentum tensor is unclear. However, the issue of the relationship between the averaged metric and the average motion of the physical objects in the corresponding spacetime geometry has received less attention. 

One natural proposal which, as far as we know, has not been considered before, is to define the averaged effective spacetime metric, at a given scale $L$, as that for which the world lines of the CoM of suitable test objects of proper size $L$ are geodesics
 --- section \ref{CoMcs} clarifies what such objects should be.
 The appealing feature of this proposal is that an effective equivalence principle is built in \textit{ab initio} for the effective metrics. On the other hand, if we take that approach too literally, we face a clear breaking of that principle, reflected in the fact that there is, in general, no universal scale-independent metric. The point, of course, is that we should consider that the ``true'' metric corresponds to the smallest possible scale, as this is dictated by the available probes. As far as we can see, in most applications to cosmological and astrophysical problems, this approach does not face any serious drawbacks, as one can never go to scales below those characteristic of, say, a grain of dust. Moreover, one should have  a clear understanding that the effective metric can only be considered as a useful auxiliary tool. 
 %A detailed analysis shows, however, that things may not be %straightforward; see the discussions in section \ref{Diss}. %The point is that, at the more fundamental level, trying to %go smallest scales, and trying to take the program as a %foundational one providing a characterization of the %emergence of a classical metric, would take us into the %quantum realm, even before getting into the full quantum %gravity realm where things are even more problematic.
 
We should also stress that the idea of a spacetime that emerges from more fundamental degrees of freedom is not new. There are many models in the literature proposing a realization of such a process \cite{emergent1,emergent2,emergent3,emergent4,emergent5,emergent6,emergent7,emergent8,emergent9,emergent10,emergent11,emergent12,emergent13,emergent14}, however, our approach is different since we look for possible fundamental obstructions in this program (for studies in the same spirit see references \cite{salecker,carlip}).

The initial exploration of the issues described above is the goal of the present manuscript, which is organized as follows: In section \ref{emergence} we list the main steps required to obtain an emergent geometrical description from the fundamental quantum theory of gravity and we specify in which step we perform our analysis. Next, in section \ref{effective geom} we deal with the program of extracting an effective geometry at a certain coarse-grained scale, from a smaller scale, focusing on those aspects  that already appear in the classical description. We also explore a number of approaches for the extraction of an effective geometry using extended but classical objects. Section \ref{quantum aspects} is devoted to further difficulties including some aspects of the problem which, we believe, could have implications in any attempt to extract a classical spacetime from a more fundamental theory of quantum gravity. Finally, in section \ref{Diss} we discuss the significance of our findings and future prospects for this line of research.

\section{The emergence of general relativity}\label{emergence}

We now frame the problem as a sequence of regimes, and of steps connecting them, providing a path  to recover GR from a fundamental quantum gravity description. 
\begin{enumerate}

\item \textbf{Full Planck-scale regime.} The fundamental spacetime degrees of freedom are characterized in terms of an appropriate quantum gravity language. These degrees of freedom, when excited in appropriate manners, correspond to what we call spacetime, which however, at this point is represented in terms of pregeometrical notions, such as the excitations of a spin network, events of the causal sets, simplices in dynamical triangulations, \textit{etc.}, and the matter degrees of freedom are characterized by secondary and subsidiary mathematical objects intrinsically tied to the former. Thus, the matter degrees of freedom must be thought of as ``attached'' to the fundamental pregeometrical degrees of freedom as a natural way to represent, at the basic level, the general idea that matter ``lives'' in spacetime. 
 
\item \textbf{Hard semi-classical gravity regime.} Spacetime geometry is characterized by a metric and the matter degrees of freedom are described in the language of quantum fields in curved spacetime. However, in this regime, the matter states are highly nonclassical and the curvature variations over the regions of interest are large, although all measures of curvature are below the Planck scale. We think of this regime as representing situations where the changes in curvature are large even over regions where the fundamental excitations of the matter fields have nontrivial support. 

\item \textbf{Soft semi-classical gravity regime.} The spacetime geometry is characterized by a metric and the matter degrees of freedom are described in the language of quantum fields in curved spacetime. Here we assume that the matter fields are in states that result in a classical-like behavior and the metric has been smoothed out by averaging over suitable regions, leading to  strong bounds on curvatures and their variations (e.g., with variations of, say, the curvature scalar, $\Delta R$, over spacetime volumes of size $L^4 $  bounded by $ |\Delta R| < 1/ L^2$).
 
\item \textbf{GR with point particles}. There is a classical description of spacetime in terms of a smooth metric and where we have a classical description of matter in terms of collection of particles moving along some specific world lines. 
 
\item \textbf{GR with fluids}. Spacetime is described in terms of a smooth metric and matter is modelled as a fluid, which describes, in an effective way, a large collection of classical particles. 
\end{enumerate}
 
It is worth noting that the regime 5 is, in fact, the regime for which GR was formulated and where the theory has been significantly tested, leading to excellent agreement between observations and predictions \cite{GRtests}. In the regime 4 we believe that GR is reliable, but, in fact, the theory has only been partially tested using test point particles (say elementary particles as gamma rays, neutrinos, and perhaps the protons and heavier nuclei constituting the high-energy cosmic rays) in the context of astrophysics \cite{AstrophysTest} and cosmology \cite{CosmolTests}, as well as simple test of Newtonian gravity in the lab with elementary particles \cite{QuantumTests}. However, as far as we know, there are no tests of the role of such elementary particles as sources of gravitation. Most GR precision tests are actually made with extended objects such as torsion pendula \cite{Adelberger} or smooth spheres as in Gravity Probe B \cite{GravProbeB}. Regarding regime 3, we can point only to the studies of inflation and the generation of primordial cosmic perturbations \cite{inflation} as examples where empirical data supports the results of the theoretical analysis. Needless to say that the regimes 1 and 2 remain, up to day, empirically unexplored.
 
According to our current theoretical understanding, we have some ideas on how to connect the different regimes. For regime 1 we have essentially no control and, although there are several promising proposals that are being actively pursued, there is no theory that can be thought of as definite, reliable, and fully workable. Thus, it is clear that the step connecting 1 to 2 is not available either. In fact, as is well known, if the regime 1 is based on a canonical approach, what one faces is a timeless theory, and the task of recovering time is nontrivial \cite{ProblemTime}.

Regime 2 is  where semiclassical gravity together with quantum field theory in curved spacetime are meant to be valid. The implicit connection between the two theories is straight forward: the quantum field theory is constructed over the spacetime metric $g_{ab}$ in the standard fashion (say, using the algebraic approach \cite{AlgebraicQFT}) and the matter fields are supposed to be characterized by a particular quantum state ${\omega} $ such that Einstein's equations hold for the renormalized expectation value of the energy-momentum tensor, ${\omega} [ T_{ab}^{Ren}]$, which acts as the gravitational source. Here one faces the issue of fully characterizing the renormalization and specifying the finite counter-terms but, otherwise, things are under control. A similar description holds for regime 3, and thus, passing from regime 2 to 3 should be straightforward. To do that, one would start with the expectation value ${\omega} [ T_{ab}^{Ren}] (x)$ on some appropriate state characterizing a situation in regime 2 and average it over spacetime regions of a suitable size to obtain a smeared object $\bar \omega[ T_{ab}^{Ren}] (x)$, which would then be used to define an effective metric $g_{ab}^{(3)}$ through some compatibility conditions. However, this part of the program is technically demanding and it is not studied here.
 
Regime 4 is where we seem to face no fundamental questions and every aspect of the theory is, in principle, under control. Free point particles follow the geodesics of the effective metric in this regime, $g_{ab}^{(4)}$, and the matter energy-momentum tensor is simply
\begin{equation}\label{EMT4}
T_{(4)}^{ab} (x) = \sum_{i} m_{i} \int d\lambda u_{i}^a u_{i}^b \delta (x-\gamma_i) , 
\end{equation}
where $ m_i$ and $u_{i}^a (\lambda) $ are the $i$th particle mass and four-velocity, respectively. For simplicity we are considering these particles to be noninteracting and we denote the corresponding geodesics by $\gamma_{i} (\lambda) $, where $\lambda $ is the corresponding parameter. Consider the step connecting the regimes 3 and 4, i.e., the transition from localized field excitations to effective point particles. This step involves caclulating  $ <\omega| T_{ab}^{Ren}|\omega> $, identifying the localized excitations, and replacing them with effective point particles of mass $ m_i$, and  world lines $\gamma_i$. Given these world lines, we would need to identify them as the geodesics of $g_{ab}^{(4)}$. That is, we would need to construct a tensor field $g_{ab}^{(4)}$ such that, for all $i$, $ u_{i}^a (\lambda) \nabla_a u_{i}^b (\lambda) =0$ where $ \nabla_a$ is the derivative operator associated with $g_{ab}^{(4)}$. 
 
Regime 5 is the one for which GR was constructed and which is used in almost any relevant application. Note that we often go back to regime 4 when considering things like planetary motion in the solar system, but we take that just as a very good approximation, and when dealing with high precision studies, we introduce corrections that take into account complexities of the finite sizes of the objects under consideration. However, the transition from regime 4 to 5 is also nontrivial. To actually achieve it, we would need, in principle, a recipe to smear the nonsmooth tensor $ T_{(4)}^{ab} $ over suitable spacetime regions, leading to a smooth averaged tensor field $T_{(5)}^{ab}$. Then, the construction of the effective metric suitable in this regime, $g_{ab}^{(5)}$ would need to ensure that the corresponding conservation equation holds. Again, in most practical examples, the input is probably not enough to completely determine $g_{ab}^{(5)}$, but it is certainly a consistency requirement for a situation where we expect GR to hold. 
 
The task at hand would be to explore some of the issues that arise in the various steps, primarily the question of geometry extraction from the behavior of matter probes, and the characterization of the energy-momentum tensor of the effective bodies from the immediate previous level. The task is clearly a formidable one, and thus, here we address it using a simplified version of the problem. The simplified analysis we carry out contemplates the passage from regime 4 to a regime 4$'$ corresponding simply to one characterized by a larger scale. The idea is that starting from regime 3 we may use different averaging scales and arrive at various versions of regime 4. We want thus to consider starting from the regime 4 tied to a certain scale $L$ and then going to the regime 4$'$ corresponding to the scale $ L' $ with $L' \gg L $. In other words, we want to consider the steps leading to the generation of the effective world lines, effective metric, and effective energy-momentum tensor of the regime 4$'$ assuming we are given these same elements at the level of the regime 4.
 It is worth noting that we do not consider the very important question of the degree to which Einstein's equations for regime 4$'$ are satisfied.

\section{Effective geometry}\label{effective geom}

We want to study the change on the geometrical quantities that can be read off from the motion of extended probes when the probing objects scale is modified. For this purpose we need to extract a world line characterizing the probe's evolution, which is to be identified as a geodesic of the effective spacetime metric. Clearly, these world lines have to be defined in a canonical and background independent way, otherwise we might introduce structures that are in conflict with the underlying diffeomorphism invariance ruining any chance of recovering GR. The most natural option available corresponds to choosing a covariant center of mass associated with the matter distribution given in terms of the classical energy-momentum tensor (or by the expectation value of the energy-momentum tensor operator in the quantum state of the matter fields). There could be alternative methods to assign world lines to extended objects not involving the center of mass, but they involve, in some way or another, similar problems. However, the method we are advocating seems to be the one that better realizes the equivalence principle in the sense of establishing, by construction, that free objects move along geodesics. 

We assume that we are presented with a background spacetime with metric $g_{ab}$ and we try to extract an effective geometry by analyzing the center of mass trajectory for some extended test objects. Observe that the basic aspects of our method are already present in the simple setting where the object's internal structure is characterized in terms of a multipolar expansion \cite{Dixon}. In fact, in the limit where the size of that object is taken to zero but the expansion is truncated at the second (dipolar) term, the evolution of the system can be described by the so-called Papapetrou equations \cite{Papapetrou} 
\begin{eqnarray}
\label{Papap1} 
T^c \nabla_c S^{ab}+T_cT^bT^d\nabla_d S^{ac}+T_cT^aT^d \nabla_d S^{bc}
&=
0
\, ,
\\
\label{Papap2} 
T^d \nabla_d(M T^a +T_b T^c\nabla_c S^{ba})-\frac{1}{2}S^{bc} T^d {R_{bcd}}^a
&=
0,
\end{eqnarray}
where $T^a$ is the unit tangent to the object's world line, $S^{ab}$ and $M$ stand, respectively, for the total spin and total mass of the object (whose definitions are given below), and $\nabla_a$ and ${R_{abc}}^d$ are the derivative operator and the Riemann curvature tensor associated to $g_{ab}$. All tensors are evaluated on the corresponding point on the object's world line. Evidently, the presence of spin and a nonconstant mass produce a deviation of the CoM trajectory from a geodesic of the background geometry. Thus, the procedure we describe above would fail to recover the true backgound geometry. Nonetheless, we can still ask if for a given class of extended objects, there is an effective geometry such that its geodesics coincide with the corresponding CoM trajectories, and which depends on the extended objects' characteristics.

\subsection{Center of mass}\label{CoMcs}

This subsection is based on the definition of the covariant (i.e., observer independent) CoM in GR originally formulated by Dixon \cite{Dixon}. This definition applies to an extended object described in terms of its energy-momentum tensor, however, for simplicity we focus in the particular case of a collection of $N$ free test massive point-like particles; all the steps we describe generalize straightforwardly to continuous matter distributions by properly changing summations to integrals. To be certain that the CoM can be defined, we assume the hypotheses of reference \cite{Beiglbock} which essentially imply that the spacetime curvature at the position of the object is small in comparison with (the inverse of the square of) the object's size, in such a way that the  object's world tube is contained in a normal convex hull $\mathcal{U}$.

Our intention is to generalize the special relativistic (SR) notion of the \emph{centroid} of a system of $N$ particles, with respect to an observer at $x_0$ with four velocity $U_0$, given by
\begin{equation}
\label{centroiddefSR}
\Xi^a(x_0,U_0)=\frac{\sum_{i=1}^N E_i(U_0) \Xi^a_i(x_0,U_0)}{\sum_{j=1}^N E_j(U_0)}
\, ,
\end{equation}
where $\Xi^a_i(x_0,U_0)$ is the $i$th particle position in the observer's frame, and $E_i(U_0)=U_0 \cdot P_i$ is the corresponding energy, as well as the covariant CoM definition obtained by applying the above centroid recipe in the rest frame of the extended object, where the total momentum has vanishing spatial components, \emph{i.e.}, it is parallel to the frame's four velocity. 
Now, there are three basic questions that we need to answer before we are able to generalize a special relativistic definition to the GR case: 
\begin{enumerate}
\item
What is to replace the SR simultaneity hyperplane, which is orthogonal to an observer's four velocity? 
\item
How can positions be treated as vectors in a general spacetime, like they are in SR?
\item
How are vectors in distinct events to be summed together, as is needed in computing, \emph{e.g.}, the total momentum of a system?
\end{enumerate}
The short answers to these are, respectively:
\begin{enumerate}
\item
Let each time-like four velocity define a simultaneity hypersurface generated by geodesics orthogonal to it.
\item
Use the (inverse of) the exponential map to give vector coordinates to all points in the region of interest.
\item 
Use parallel transport along the unique geodesic connecting a general point to some fixed base point, to bring all vectors in the same tangent space there, where they can be meaningfully summed.
\end{enumerate} 
We now elaborate on these answers:
let $x_0$ be a point in $\mathcal{U}$ and $U_0^a$ a time-like, future-directed unit vector tangent to $\mathcal{U}$ at $x_0$, \emph{i.e.}, belonging to the tangent space $V_{x_0}$. We first construct the simultaneity surface with respect to $U_0^a$, denoted by $\Sigma(x_0,U_0)$, which is given by all events that can be reached from $x_0$ by a geodesic passing through $x_0$ and whose tangent, at this point, is orthogonal to $U_0^a$. Note that $\Sigma(x_0,U_0)$ may, in general,  not be a hyper-surface, but the hypotheses we make are such that, its restriction to $\mathcal{U}$ is indeed a space-like hyper-surface. Denote by $y_{i}(x_0,U_0)$ the point where the $i$th particle world line intersects $\Sigma(x_0,U_0)$; since we have assumed that $\mathcal{U}$ is a normal convex hull, there is only one geodesic connecting $y_{i}(x_0,U_0)$ and $x_0$ for every particle. This allows to unequivocally find the vectors $\Xi_{i}^a(x_0,U_0)\in V_{x_0}$ that ``point in the direction of the particles'', at the instant defined by $U^a_0$, by imposing two conditions:
(i) The $\Xi_{i}^a(x_0,U_0)$ are tangent to $\Gamma(x_0,U_0)$, namely,
$U_0 \cdot \Xi_{i}(x_0,U_0)=0$,
where $X \cdot Y \equiv g_{ab}X^aY^b$, and (ii)
the geodesic that passes through $x_0$ with tangent $\Xi_{i}^a(x_0,U_0)$ intersects the $i$th particle world line at an affine distance $1$, \emph{i.e.}, in terms of the exponential map, 
\begin{equation} 
\label{cond 2}
\exp_{x_0}\left[\Xi_{i}^a(x_0,U_0)\right]=y_{i}(x_0,U_0).
\end{equation}

We also parallel transport  the momentum $P^a_i$ of the $i$th particle from $y_{i}(x_0,U_0)$ to $x_0$, along the unique geodesic joining these two points, and denote the result of this operation by $\tilde{P}^a_{i}(x_0,U_0)$. The total momentum at $x_0$, with respect to $U_0^a$, is then defined as
\begin{equation}
\label{PdefGR}
P^a(x_0,U_0)= \sum_{i=1}^N\tilde{P}^a_{i}(x_0,U_0)
\, .
\end{equation}
Note that the  sum in the right hand side is well defined, as all $P^a_i$ belong to $V_{x_0}$, and that $P^a(x_0,U_0)$, thus defined, is a time-like vector.
Finally, observer \emph{and} position-dependent energies $E_i(x_0,U_0)$ may be assigned to the particles by 
\begin{equation}
\label{Eidef}
E_i(x_0,U_0) = U_0 \cdot \tilde{P}_i(x_0,U_0)
\, .
\end{equation}
We have now at our disposal all the elements necessary to define the GR version of the centroid, 
\begin{equation}
\label{XidefGR}
\Xi^a(x_0,U_0)
=
\frac{\sum_{i=1}^N E_i(x_0,U_0) \Xi^a_i(x_0,U_0)}{\sum_{j=1}^N E_j(x_0,U_0)}
\, ,
\end{equation}
where we use the same symbol for the GR centroid as in the SR case, given that the latter is a special instance of the former.

Finally, to arrive at a covariant GR CoM recipe, we need to define a ``rest frame'' for our extended object. For some fixed base point $x_0$, we look for that special frame velocity $U_0(x_0)$ that results in the total momentum of~(\ref{PdefGR}) being parallel to $U_0(x_0)$, 
\begin{equation}
\label{U0def}
P^a(x_0,U^b(x_0))= \mu \,  U^a(x_0)
\, ,
\quad \mu \in \mathbb{R}
\, .
\end{equation}
It has been shown \cite{Beiglbock} that for a given extended object and under the work hypotheses, $U^a(x_0)$ exists and it is unique. This is then taken to uniquely define the rest frame of the extended object, with origin at the event $x_0$. We define the $x_0$-centered rest-frame momentum of the $i$th particle and total momentum of the system, as
\begin{eqnarray}
\label{Pitdef}
\tilde{P}_{i}^a(x_0)
&=&
\tilde{P}^a_{i}(x_0,U^b(x_0)),
\\
\label{Ptotdef}
P^a(x_0)
&=&
P^a(x_0,U^b(x_0)).
\end{eqnarray}
Similarly, the $x_0$-centered rest-frame energy of the $i$th particle is 
\begin{equation} 
\label{def E}
E_{i}(x_0)=U(x_0) \cdot \tilde{P}_i(x_0)
\, ,
\end{equation}
so that a reasonable attempt at a GR version of the covariant SR CoM recipe is
\begin{equation}
\label{CoMGRcov1}
\Xi^a(x_0)
=
\frac{\sum_{i=1}^N E_i(x_0) \Xi^a_i(x_0)}{\sum_{j=1}^N E_j(x_0)}
\, ,
\end{equation}
(where $\Xi^a_i(x_0)=\Xi^a_i(x_0,U^b(x_0))$) with the understanding that the actual CoM is the exponential map of the above vector (see Fig.~\ref{CMconventions}).
%%%%%%%%%%%%%%%%%% FIGURE
\setlength{\figurewidth}{.8\textwidth}
%\begin{floatingfigure}{.93\figurewidth}
\begin{figure}[h]
%\rule{0mm}{.675\figurewidth}
\centerline{%
\begin{pspicture}(-.4\figurewidth,-0.2\figurewidth)%
                 (.4\figurewidth,.5\figurewidth)
\setlength{\unitlength}{.25\figurewidth}
\psset{xunit=.25\figurewidth,yunit=.25\figurewidth,arrowsize=1.5pt
3}
%\psgrid[subgriddiv=10,griddots=5,gridlabels=8pt]
%%%%%%%% Left particle worldline
\pscurve[linewidth=.3mm]{-}%
(-1.5,-.7)(-1.4,.4)(-1.0,1.8)
%%%%%%%% CoM worldline
\pscurve[linewidth=.5mm,linecolor=gray]{-}%
(-.4,-.7)(-.3,.4)(.0,1.8)
%%%%%%%% Right particle worldline
\pscurve[linewidth=.3mm]{-}%
(1.0,-.7)(1.3,.4)(1.3,1.8)
%%%%%%%% Simultaneity hypersurface \Sigma
\pscurve[linewidth=.5mm,linecolor=gray,linestyle=dashed]{-}%
(-1.7,0.08)(0.0,0.34)(1.6,0.13)
%%%%%%%% Puts
%%% Points
\psdot[dotsize=1.2ex](-1.44,.135)        % Particle 1 dot
\psdot[dotsize=1.2ex](1.265,.193)          % Particle 2 dot
\psdot[dotsize=1.2ex](-.31,.33)             % x_0 dot
\psdot[dotsize=1.2ex](.4,.315)             % CoM dot
%%%%%%%%% \Xi's - arrows
\psline[linewidth=.6mm,linecolor=black]{->}% \Xi_1 arrow
(-.31,.33)(-1.4,.24)
\psline[linewidth=.6mm,linecolor=black]{->}% \Xi_2  arrow
(-.31,.33)(1.28,.44)
\psline[linewidth=.6mm,linecolor=black]{->}% \Xi_CoM  arrow
(-.31,.33)(.384,.38)
%%%%%%%%% Momenta - arrows
\psline[linewidth=.4mm,linecolor=black]{->}% P_1 arrow
(-1.44,.14)(-1.37,.84)
\psline[linewidth=.4mm,linecolor=black]{->}% P_2  arrow
(1.265,.193)(1.45,1.1)
\psline[linewidth=.6mm,linecolor=black]{->}% P_CoM  arrow
(-.31,.33)(-.15,1.8)
\psline[linewidth=.4mm,linecolor=black]{->}% \tilde{P}_1  arrow
(-.31,.33)(-.4,1.0)
\psline[linewidth=.4mm,linecolor=black]{->}% \tilde{P}_2  arrow
(-.31,.33)(.05,1.3)
%%%%%%%%% CoM pointing  arrow
\psline[linewidth=.2mm,linecolor=black]{->}% CoM-pointing  arrow
(.3,.1)(.38,.28)
%%%%%%%%% Particles - labels
\put(-1.27,.02){\makebox[0cm][r]{$\pi_1$}}  % \pi_1 label
\put(1.09,.1){\makebox[0cm][l]{$\pi_2$}}  % \pi_2 label
%%%%%%%%% Positions - labels
\put(-1.31,.35){\makebox[0cm][l]{$\Xi_1(x_0)$}}  % \Xi_1 label
\put(1.15,.54){\makebox[0cm][r]{$\Xi_2(x_0)$}}  % \Xi_2 label
\put(.28,.47){\makebox[0cm][r]{$\Xi(x_0)$}}  % \Xi_CoM label
%%%%%%%%% Momenta - labels
\put(-1.36,.87){\makebox[0cm][r]{$P_1(x_0)$}}  % P1 label
\put(1.48,1.1){\makebox[0cm][l]{$P_2(x_0)$}}  % P2 label
\put(-.38,1.85){\makebox[0cm][l]{$P(x_0)$}}  % P_CoM label
\put(-.43,1.03){\makebox[0cm][r]{$\tilde{P}_1(x_0)$}}  % \tilde{P}_1 label
\put(.02,1.34){\makebox[0cm][l]{$\tilde{P}_2(x_0)$}}  % \tilde{P}_2 label
%%%%%%%%% Hypersurface \Sigma - label
\put(1.58,.0){\makebox[0cm][r]{$\Sigma$}}  % \Sigma label
%%%%%%%%% Hypersurface x_0 - label
\put(-.15,.21){\makebox[0cm][r]{$x_0$}}  % x_0 label
%%%%%%%%% CoM - label
\put(.2,.0){\makebox[0cm][l]{$\exp[\Xi(x_0)]$}}  % CoM label
\end{pspicture}%
}
\caption{%
Schematics of Dixon's $x_0$-centered CoM definition for a two-particle system. For \emph{the} CoM $x_0$ should be such that $\Xi$ vanish.
}
\label{CMconventions}
\end{figure}
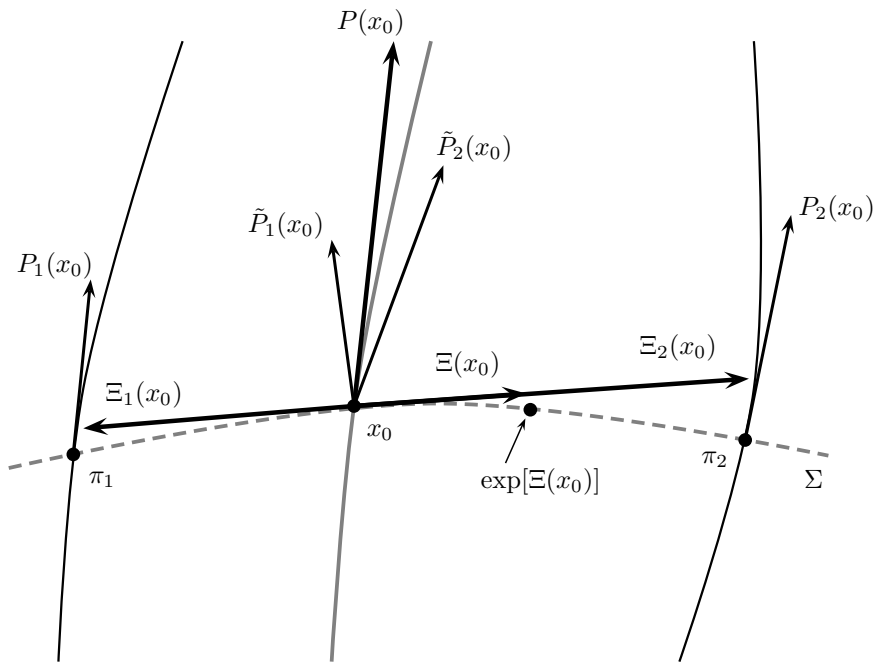
%\end{floatingfigure}
%%%%%%%%%%%%%%%%%% FIGURE
We are almost done, but not quite at the end yet, as the above definition depends on the base point $x_0$. We emphasize that this is \emph{not} obvious from~(\ref{CoMGRcov1}): even an $x_0$-independent CoM would be described by an $x_0$-dependent normal coordinate, as moving, say, $x_0$ further away from the CoM worldline would result in a larger vectorial coordinate, the exponential map of which would land one on exactly the same worldline as before. The non-obvious fact is that the vectorial normal coordinate $\Xi^a(x_0)$ in~(\ref{CoMGRcov1}) fails to exactly compensate for the movement of $x_0$, as described above, so that the CoM worldlines computed in frames with distinct origins do not, in general, coincide. The last element in our series of definitions is obtained by finding those special points  $x_0=\bar{x}$ such that $\Xi^a(\bar{x})=0$ --- the locus of the $\bar{x}$ gives \emph{the} CoM worldline, independent of any observer characteristics.
It is proven in~\cite{Beiglbock} that the above CoM worldline is a unique continuous and everywhere time-like curve.

Having at our disposal the vector fields $\Xi_i^a(x)$, $\tilde{P}_i^a(x)$, we define a total angular momentum tensor field via
\begin{equation} 
\label{def J}
J^{ab}(x)
= 
\sum_{i=1}^N 
\Xi_{i}^{a}(x) \tilde{P}^{b}_{i}(x) - \Xi_{i}^{b}(x) \tilde{P}^{a}_{i}(x)
\, .
\end{equation}
Contracting this with the total momentum field gives a new vector field, $W^a(x)=g_{bc}(x)P^b(x) J^{ca}(x)$, which can be shown to vanish on the CoM worldline, $W(\bar{x})=0$.
This CoM property was crucial in the first covariant definition in curved spacetimes \cite{Dixon} and it is sometimes used to motivate alternative CoM definitions \cite{Madore}. The total spin $ S^{ab}$ of the extended object is then defined as the total angular momentum with respect to the CoM, that is, $S^{ab}=J^{ab}(\bar{x})$. In addition, the total mass is defined as the total momentum norm, evaluated on  the CoM world line, that is
\begin{equation}
M=\sqrt{g_{ab}(\bar{x})P^a(\bar{x})P^b(\bar{x})}.
\end{equation}
It turns out that, in the limit when the object size tends to zero, the CoM world line is obtained from the Papapetrou equations (\ref{Papap1})-(\ref{Papap2}), and thus, it generally fails to be a geodesic of the background metric. On the other hand, in flat spacetime, the objects we consider as probes always follow geodesics \cite{Pryce}. We now  examine two different methods to extract  information about the effective geometry by identifying the CoM worldline with the geodesics of an effective metric.
 
\subsection{Effective connection}\label{Eff con}

In this section we assume that we can monitor the center of mass position of a given class of extended objects, and thus, we have the manifold crisscrossed by world lines. From these world lines we seek to extract some notions of an effective geometry. 
In particular, to extract an effective connection we use the property that the vector field $W^a(x)$ vanishes along the CoM worldline. Let $T^a$ be the unit tangent vector of the CoM worldline, then the second derivative of $W^a$ along this line is
\begin{eqnarray}
\label{EffCon}
0
&=&
T^c \nabla_c(T^b \nabla_b W^a)\vert_{\bar{x}}
\nonumber
\\
&=&
\dot{T}^b \partial_b W^a\vert_{\bar{x}}
+
T^b T^c \left(\partial_c\partial_b W^a
+
2 \Gamma^a_{bd}(Z) \partial_c W^d\right)\vert_{\bar{x}}
\, .
\label{geom efect}
\end{eqnarray}
The above equation may be cast into the form \cite{Yuri}
\begin{equation}
\label{def Gamma tilde}
0=\dot{T}^m + T^r T^s \gamma_{rs}^m(\bar{x})
\, ,
\end{equation}
where  $\gamma_{rs}^m$ are the effective connection components.

A detailed examination of this method reveals a serious obstacle: from equation (\ref{EffCon}) one can only read the effective connection components along the CoM world line and not in the perpendicular directions. To extract these latter components we would have to probe the same spacetime region in other directions with precisely the \emph{same} extended object. However, in a general spacetime it is not possible to ensure that two objects in different regions are identical. This issue seems to be related with the well-known obstacles for defining rigid bodies in GR. Therefore, in general, these effective connection components would depend on the details of the extended objects that are used, calling into question the possibility of recovering an object-independent spacetime geometry.

\subsection{Effective curvature}\label{eff curv}

Another approach that offers more direct and relevant information about the effective geometry is to read off directly the effective sectional curvature. Recall that this curvature can be measured through the relative acceleration of infinitesimally close geodesics. We can then define the effective sectional curvature as codifying the relative acceleration of the CoM world lines of neighboring free extended objects.

More concretely, we use a uniparametric family of extended objects where the parameter $\kappa$ selects the extended object. Also, we denote by $\bar{x}(\kappa, \lambda)$ the CoM world line of the object $\kappa$, which is, in turn, parameterized by $\lambda$, and by $T^a$ the unit tangent of the CoM world line of each object, that is $ T^a \propto (\partial \bar{x}/\partial \lambda)^a$. Then, the deviation vector of neighboring geodesics is
\begin{equation}
X^a=\left(\frac{\partial \bar{x}}{\partial \kappa}\right)^a+b T^a,
\end{equation}
where $b$ is fixed by the condition that $X^a$ be orthogonal to $T^a$. Moreover, the relative acceleration, $a^a=T^c \nabla_c (T^b \nabla_b X^a)$, is linked to the effective curvature tensor in the probed region, ${r_{abc}}^d$, via
\begin{equation}
\label{desv}
a^a=-{r_{bcd}}^a T^b X^c T^d.
\end{equation}
A series of calculations, using this approach, have been carried out in~\cite{OurPaper}, in the case where the underlying spacetime has constant curvature, and considering, for simplicity, the centroid of theprobe rather than the actual CoM. We found that, in general, the effective geometry depends on the details of the extended object playing the role of the free particle. In fact, in extreme cases, e.g., if the object is large enough or hot enough (i.e., if it has sufficient internal energy), the effective sectional curvature can be of opposite sign as compared to the one of the underlying metric. This, by itself, seems to cast serious doubts on the possibility of recovering an object-independent effective spacetime geometry. In addition, in reference \cite{OurPaper} it is shown that further complications arise when attempting to fulfill the program outlined above. First, the effective sectional curvature depends on the relative acceleration of the center of mass world lines of two extended objects. Therefore, the issue of whether these objects are identical is also present in this case. In reference \cite{OurPaper}  such problems were avoided by using the background spacetime symmetry, but it is clearly impossible to bypass this issue in general. Second, the effective curvature can be read off only in the plane defined by the vectors $X^a$ and $T^a$; the other components of ${r_{abc}}^d$ cannot be defined without considering additional objects, and of course, that brings back the issue of defining identical extended objects in curved spacetimes. Remarkably, all these complications do not appear in flat spacetime, because, in that case, the CoM always follows a geodesic. Note that, as discussed in that work, the main features and problems encountered in reference \cite{OurPaper} are expected to be present when using the CoM, instead of the centroid, of the probe.

The fact that, in this approach, the dependence of the CoM world line on the object's internal structure goes beyond that specified by the total spin of the extended object, involving higher multipole momenta, could, in fact, become an obstacle for defining canonical probes with which to carry out our program. This issue needs further investigation, which is left out of the present work.

\section{Additional aspects}\label{quantum aspects}

\subsection{An overlooked failure: nonassociativity}
\label{nonassoc}

Despite its long history, the CoM recipe presented in section~\ref{CoMcs} suffers from a consistently overlooked shortcoming: in a sense we make precise shortly, it fails to be associative. Associativity is a property of some sort of product, so, it is not surprising that our more precise take on this concept, which we now undertake, starts with defining a product: given two point particles $\pi_A$ and $\pi_B$, call $T_A$, $T_B$, their respective energy-momentum  tensors, that have support on the particles' world lines. A center of mass recipe, in its simplest form, should accept $T_A$, $T_B$ as input, and produce an effective point particle, call it $\pi_{AB}$, described by the energy-momentum tensor $T_{AB}$, with support on the world line of $\pi_{AB}$. Thus, we may introduce a $*$-product between point-particle energy-momentum tensors, and write for the above operation $T_{AB}=T_A * T_B$. We certainly want the operation to be commutative, given that there is no intrinsic way to define the order of the factors, so that we demand $T_A * T_B =T_B * T_A$. 

What other properties are needed for the product $*$ to be satisfactory? A basic question emerges naturally when considering $n$-particle systems: is the above binary $*$-product sufficient to determine the center of mass by repeated application, or does such a system require a new, $n$-ary product, that cannot be reduced to binary operations? The former option is undoubtedly preferable, on both aesthetic and practical grounds. But if this is going to be the case, the $*$-product has better be associative, so as to define a unique result, regardless of the bracketing of the factors used. For example, for a three-particle system, one could compute $T_{(AB)C}=(T_A*T_B)*T_C$ and, in a similar notation,  $T_{A(BC)}$ or $T_{(AC)B}$ --- it would certainly be desirable that all these results coincide. 

How does the Dixon center of mass definition perform with respect to the above criteria? First, it is easy to see that the application of this recipe to a two-particle system defines a commutative but non-associative $*$-product. Furthermore, its application to an $n$-particle system, with $n>2$, gives rise to an $n$-ary product that is not related in any discernible way to repeated applications of the $*$-product. Thus, for example, if the center of mass of an $n$-particle system $\{\pi_1,\ldots,\pi_n\}$ has been computed \emph{\`a la} Dixon, and an additional particle is subsequently included in the system, the calculation of the center of mass of the resulting $(n+1)$-particle system has to start from scratch --- there is simply not enough information in the energy-momentum tensor $T_{1\ldots n}$ to adequately represent the $n$-particle system in the $(n+1)$-particle calculation.

A similar problem occurs in flat spacetime when the effective intrinsic spin of the objects is not taken into account \cite{Chryssomalakos:2009}. The recipe considers point like particles which have no intrinsic spin, and  represents an extended object by its center of mass. When the spin of the composed system is not taken into the account, the resulting recipe fails to be associative. The proposal of reference \cite{Chryssomalakos:2009} is to include the spin of every component of the extended object when calculating its total angular momentum, and, with this modification, the CoM definition in flat spacetime becomes associative. However, this simple solution does not cure all the causes that make the GR CoM nonassociative.

If Dixon's CoM definition fails to be associative, what particular characteristic of the system causes the failure? To answer this question fully, in the general case, is a rather complicated matter --- what we attempt here is a simplified treatment, aiming to give us a hint of the nature of the phenomenon. We consider a three-particle system $\{\pi_1,\pi_2,\pi_3\}$, and ask under what circumstances is the center of mass description of the system at its most accurate? A true point particle has a sharp position, and a sharp momentum, corresponding to a single point in phase space. A system of particles will be well approximated by its CoM if its extension both in position and momentum space is small. For the position, the scale is set by the radius of curvature of spacetime; for the momentum, by the masses of the particles. Assuming both of these scales to be of order unity, we expect simplifications to occur, when the system may be fit inside a parallelepiped of sides $\epsilon$, $\eta \ll 1$ in phase space, along position and momentum axes, respectively --- the particles have similar masses, and travel close to each other, with almost parallel four-momenta. Under these assumptions, it is natural to expand all quantities of interest in a double power series in $\epsilon$ and $\eta$ --- one could hope that, for example, the leading term(s) in such an expansion would be both informative and computationally accessible. The particular quantity we compute along these lines is the difference $\delta \Xi_{(12)3}$ in the Dixon CoM normal coordinates for the above three-particle system,
%%%%%%%%%%%%%%%%%% FIGURE
\setlength{\figurewidth}{.8\textwidth}
%\begin{floatingfigure}{.93\figurewidth}
\begin{figure}[h]
%\rule{0mm}{.675\figurewidth}
\centerline{%
\begin{pspicture}(-.4\figurewidth,-0.2\figurewidth)%
                 (.4\figurewidth,.5\figurewidth)
\setlength{\unitlength}{.25\figurewidth}
\psset{xunit=.25\figurewidth,yunit=.25\figurewidth,arrowsize=1.5pt
3}
%\psgrid[subgriddiv=10,griddots=5,gridlabels=8pt]
%%%%%%%% Side 12
\psline[linewidth=.3mm]{-}%
(-.8,-.4)(.0,.8)
%%%%%%%% Side 13
\psline[linewidth=.3mm]{-}%
(-.8,-.4)(1.1,-.2)
%%%%%%%% Side 23
\psline[linewidth=.3mm]{-}%
(.0,.8)(1.1,-.2)
%%%%%%%% Segment 1A
\psline[linewidth=.8mm,linecolor=lightgray]{-}%
(-.8,-.4)(-.4,.2)
%%%%%%%% Segment 2A
%\psline[linewidth=1mm,linecolor=lightgray]{-}%
%(.0,.8)(-.4,.2)
%%%%%%%% Segment A\calO
\psline[linewidth=.8mm,linecolor=lightgray]{-}%
(-.4,.2)(.1,.067)
%%%%%%%% Segments 3\calO
%\psline[linewidth=1mm,linecolor=gray]{-}%
%(1.1,-.2)(.1,.067)
%\psline[linewidth=.7mm,linecolor=lightgray,linestyle=dashed]{-}%
%(1.1,-.2)(.1,.067)
%%%%%%%% Segment 1\calO
\psline[linewidth=.7mm,linecolor=gray,linestyle=dashed]{-}%
(-.8,-.4)(.1,.067)
%%%%%%%% Segment 2\calO
%\psline[linewidth=.7mm,linecolor=gray,linestyle=dashed]{-}%
%(.0,.8)(.1,.067)
%%%%%%%% Puts
%%% Point letters/numbers/dots
\put(-.87,-.43){\makebox[0cm][r]{$1$}}   %  1
\put(-.07,.85){\makebox[0cm][c]{$2$}}  %  2
\put(1.15,-.25){\makebox[0cm][l]{$3$}}  %  3
%%% Points
\psdot[dotsize=1ex](-.4,.2)            % A
\put(-.5,.2){\makebox[0cm][r]{$A$}}
\psdot[dotsize=1ex](.55,.3)              %B
\put(.58,.33){\makebox[0cm][l]{$B$}}
\psdot[dotsize=1ex](.165,-.3)             %C
\put(.169,-.47){\makebox[0cm][r]{$C$}}
\psdot[dotsize=1ex](.1,.067)            % \calO
\put(.05,-.11){\makebox[0cm][l]{$\calO$}}
%%%%%%%%% Momenta - arrows
\psline[linewidth=.6mm,linecolor=black]{->}% momentum P1 arrow
(-.8,-.4)(-.7,.2)
\psline[linewidth=.6mm,linecolor=black]{->}% momentum P2 arrow
(.0,.8)(.1,1.4)
\psline[linewidth=.6mm,linecolor=black]{->}% momentum P3 arrow
(1.1,-.2)(1.2,.4)
\psline[linewidth=.6mm,linecolor=black]{->}% momentum P arrow
(.1,.067)(.4,1.867)
\psline[linewidth=.6mm,linecolor=black]{->}% momentum \delta P arrow
(.4,1.867)(.2,1.85)
\psline[linewidth=.6mm,linecolor=black]{->}% momentum \tilde{P} arrow
(.1,.067)(.2,.667)
\psline[linewidth=.6mm,linecolor=black]{->}% momentum \tilde{P}_1 arrow
(.1,.067)(.05,.67)
%%%%%%%%% Momenta - labels
\put(-.8,.1){\makebox[0cm][r]{$P_1$}}  % P1 label
\put(-.14,1.3){\makebox[0cm][l]{$P_2$}}  % P2 label
\put(1.25,.3){\makebox[0cm][l]{$P_3$}}  % P3 label
\put(.45,1.767){\makebox[0cm][l]{$P$}}  % P label
\put(.15,1.83){\makebox[0cm][r]{$\delta P$}}  % \delta P label
\put(.2,.34){\makebox[0cm][l]{$\tilde{P}$}}  % \tilde{P} label
\put(.05,.3){\makebox[0cm][r]{$\tilde{P}_1^{1A\cal{O}}$}}  % \tilde{P} label
\end{pspicture}%
}
\caption{%
Failure of associativity in the CoM calculation of a three-particle system (see text). 
}
\label{CoM123_fig}
\end{figure}
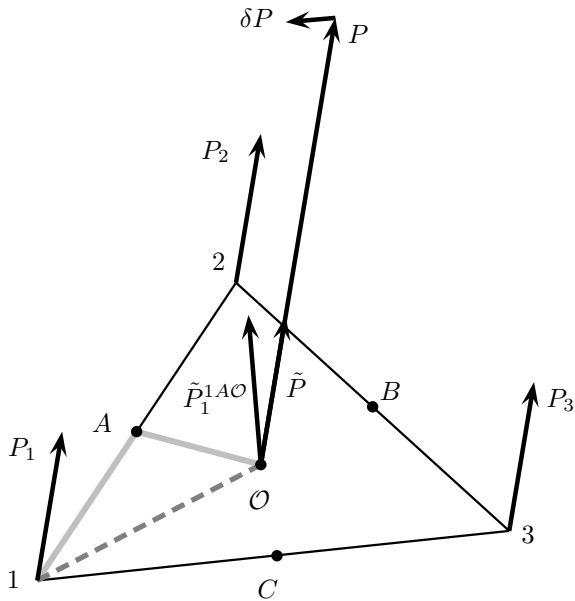
%\end{floatingfigure}
%%%%%%%%%%%%%%%%%% FIGURE
when computed in two different ways: first, we put the origin of our normal coordinate system on the 123-CoM worldline, so that the CoM normal coordinate $\Xi_{123}$ vanishes.  We then apply the recipe  to the $(12)$ subsystem, composing the result with particle 3, to get to $\Xi_{(12)3}=\Xi_{123}+\delta \Xi_{(12)3}=\delta \Xi_{(12)3}$. It should be clear, from our definition above of \emph{the} CoM of a system, that $\Xi_{(12)3}$, when exponentiated, gives a good approximation for the $(12)3$-CoM worldline, but, strictly speaking, it is not the exact result --- the latter would be found if, instead, we moved the origin of our frame so as to make $\Xi_{(12)3}$ vanish. The two approaches though coincide in the order of perturbative calculation we work with. 
 
We identify now the dominant term in $\delta \Xi_{(12)3}$. When computing $X_{123}$, the four-momenta $P_i$ are parallel-transported to the center of mass $\calO$ of the system along the geodesics connecting the vertices of the triangle 123 in figure \ref{CoM123_fig} with the center of mass $\calO$, and are summed there to give $P_{123}$. Computing $\Xi_{(12)3}$, on the other hand, involves parallel-transporting $P_1$, $P_2$ to the midpoint $A$ of $12$, along the segments $1A$, $2A$, respectively, and then transporting their sum to $\calO$, along $A\calO$, where it is added to the transported $P_3$ to give $P_{(12)3}$. The difference between the total momenta $P_{(12)3}$ and $P_{123}$ thus computed, clearly involves the holonomy of the parallel transport of $P_1$, $P_2$ along the closed circuits $1$-$A$-$\calO$-$1$ and $2$-$A$-$\calO$-$2$ respectively. Since the linear size of these two triangular circuits is of order $\epsilon$, the above holonomy is quadratic in $\epsilon$, and some detailed argumentation, that we will not invoke explicitly here, implies that this will fix  the order of the leading term in $\delta \Xi_{(12)3}$ to $\calO(\epsilon^3)$. Accordingly, we may set, for the moment,  $\eta=0$ in our calculation, which means that the parallel transported four-momenta $\tilde{P}_i$, $i=1,2,3$ (the tilde reminds the parallel transport) all coincide at $\calO$, 
\begin{equation}
\label{Ptdef}
\tilde{P}_i=\tilde{P}=E_0 U_{123}
\, ,
\quad
i=1,2,3
\, ,
\end{equation}
where $U_{123}^a=P_{123}^a/|P_{123}|$ is the 123-CoM four-velocity (we denote by $|v|\equiv \sqrt{v \cdot v}$ the norm of a time-like four vector).
 In practice then we decide what the common four-momentum $\tilde{P}$ is at $\calO$ for all three particles, and then parallel-transport it back to the vertices of the triangle, along the geodesics that connect them to $\calO$, to find what the various $P_i$ should be in order for $\eta$ to vanish. The total four-momentum, under these circumstances, is given by $P_{123}=3\tilde{P}=3E_0 U_{123}$, while the energies $E_i$ are all equal, 
\begin{equation}
\label{Eicalc}
E_i=\tilde{P}_i \cdot U_{123} =|\tilde{P}|=E_0
\, .
\end{equation} 
When computing the $(12)3$-CoM, the fact that $P_{(12)3}$ is slightly tilted with respect to $P_{123}$, implies that the simultaneity hypersurface $\Sigma_{(12)3}$, orthogonal to $P_{(12)3}$, will also be tilted with respect to $\Sigma_{123}$. The intersection point of $\Sigma_{(12)3}$ with the world line of particle $i$ will then be different from that of $\Sigma_{123}$, which will affect its vectorial coordinate $\Xi_i$. To simplify our notation, we distinguish  by a prime all quantities relevant to the $(12)3$-computation, while unprimed quantities refer to the $123$ one --- in particular, $P_{123} \rightarrow P$, and $P_{(12)3} \rightarrow P'$. Then $\delta \Xi_i = \Xi'_i-\Xi_i=\lambda P_i$, since $P_i$ is tangent to the particle $i$ world line. Orthogonality of the $\Xi$'s to the total momentum implies
\begin{equation}
\label{Xiortho}
0=\Xi'_i \cdot P'=(\Xi_i+\lambda P_i)\cdot (P+\delta P) \Rightarrow \lambda=-\frac{\Xi_i \cdot \delta P}{P_i \cdot P}
\, ,
\end{equation}
\emph{i.e.}, 
\begin{equation}
\label{Xi123}
\delta \Xi_i = \Xi_{i \, (12)3}-\Xi_{i \, 123}=- \frac{\Xi_i \cdot \delta P}{P_i \cdot P} P_i = -\frac{\Xi_i \cdot \delta P}{3E_0^2}P_i
\, .
\end{equation}
In the above expression, $\delta P \equiv P_{(12)3}-P_{123}$ is given by
\begin{align*}
\delta P
&=
\tilde{P}_{1 }^{1A\calO}
+
\tilde{P}_{2}^{2A\calO}
+
\tilde{P}_{3}^{3\calO}
- 
\tilde{P}_{1}^{1\calO}
-
\tilde{P}_{2}^{2\calO}
-
\tilde{P}_{3}^{3\calO}
\\
&=
\frac{1}{2} R_\calO(\vv{1A},\vv{1\calO})(P_1)
+
\frac{1}{2} R_\calO(\vv{2A},\vv{2\calO})(P_2) 
\, ,
\end{align*}
with $\tilde{P}_1^{1A\calO}$ denoting the parallel transport of $P_1$ along $1A\calO$, \emph{etc.}, while $R_\calO$ is the Riemann tensor at $\calO$, and $R_\calO(X,Y)(Z)^m \equiv R_{a b c}^{\phantom{a b c}m} X^a Y^b Z^c$. Noting that the matrix $R(X,Y)$ only depends on the plane of $X$, $Y$, and the (signed) area of the parallelogram they span, we may infer from the above expression that
\begin{equation}
\label{deltaP2}
\delta P=\frac{1}{12} R_\calO(X,Y)(P_1+P_2) \equiv \frac{1}{12} R_\triangle(P_1+P_2)
\, ,
\end{equation}
where $X \equiv \vv{31}$, $Y\equiv \vv{32}$ are sides of the triangle, and
 we used the fact that, due to our particular  simplifying assumptions, $\calO$ is the barycenter of the $123$-triangle (the last equality defines a convenient notation for the matrix $R_\calO(X,Y) \equiv R_\triangle$). One last result that will be needed for the calculation of $\delta \Xi$ is the change in the energy of the $i$-th particle, $\delta E_i = E'_i-E_i$. We compute 
\begin{align*}
\delta E_i
&=
\delta \left(
 \frac{\tilde{P}_i \cdot P}{|P|}
\right)
\\
&=
\frac{\delta \tilde{P}_i \cdot P}{|P|}
+
\frac{\tilde{P}_i \cdot  \delta P}{|P|}
-
\frac{\tilde{P}_i \cdot P}{|P|^3} P \cdot \delta P
\\
&=
\calO(\epsilon^2 \eta)
\, ,
\end{align*}
the last equality following from the fact that the holonomy $\delta X$ of the parallel transport of a vector $X$ along a closed circuit is orthogonal to $X$ (this is the geometrical content of the antisymmetry of $R_{a b c d}$ in its last two indices), and also  that $P \parallel \tilde{P}_i$. Note that in using this result for $\delta E_i$ in what follows, we will maintain the second term of the middle line, even though it does not contribute to the corresponding perturbative order. We conclude that individual particle energies do not change, to this order, when switching from $123$ to $(12)3$, and, hence, neither does the total energy $E$. We may now compute the change $\delta \Xi$ in the CoM vectorial coordinates,
\begin{align*}
\delta \Xi^m
&=
\delta \left(
E^{-1}\sum_i E_i \Xi_i^m 
\right)
\\
&=
E^{-1} \sum_i
\delta E_i \, \Xi_i^m + E_i \, \delta \Xi_i^m
\\
&=
E^{-2} \sum_i
 (P_i \cdot \delta P)\Xi_i^m - (\Xi_i \cdot \delta P)P_i^m
\\
&=
E^{-2} S^{m n} \delta P_{n}
\, ,
\end{align*}
where, in this order of our perturbative expansion, the tildes may be removed, and we have introduced the spin $S$ of the system, $S^{m n}=\sum_i \Xi_i^m P_i^n-\Xi_i^n P_i^m$ and the total energy $E=\sum_i E_i=|P|$. Combining with~(\ref{deltaP2}) we arrive at
\begin{equation}
\label{deltaXi3}
\delta \Xi^m = \frac{1}{12} E^{-2} S^{m n} R_{a b c n} X^a Y^b (P_1+P_2)^c
\, .
\end{equation}
A more symmetrized expression results if we average the above result over all possible ways of computing the CoM,
\begin{equation}
\label{deltaXiav}
\overline{\delta \Xi}^m =\frac{1}{18} E^{-2} S^{mn} R_{abcn} X^a Y^b P^c
\, ,
\end{equation}
where  $\overline{\delta \Xi}^m=\frac{1}{3}( \delta \Xi_{(12)3}^m+\delta\Xi_{1(23)}^m+\delta\Xi_{2(13)}^m)$ is essentially the CoM coordinate of the three CoMs, that still deviates from $\Xi_{123}^m$. Thus, the leading term responsible for the failure of associativity of the Dixon CoM definition involves the spin of the system interacting with the curvature of spacetime, a result certainly reminiscent of Papapetrou's equations. That the above result is cubic in $\epsilon$ is evident in that $X$, $Y$ are both of order $\epsilon$, as is $S^{mn}$, which involves $\Xi^a_i$ linearly.

Eq.~(\ref{deltaXiav}) may look appealing in its simplicity, but fails to deliver in one important sense: with our simplifying assumptions, the spatial part of 
$\overline{\delta\Xi}^m$ is identically zero. This is due to the fact that, for $\eta=0$, the constituent particles of our probe are at rest in the rest frame, and so the spatial part of $S^{mn}$ vanishes. All we get from~(\ref{deltaXiav}) then is the change in the time coordinate of the CoM, which is not exactly what our initial motivation was. To compute the change in the CoM position along the simultaneity hypersurface $\Sigma$, we have to push the above calculation to the $\calO(\epsilon^3 \eta)$ regime, by assuming small, $\calO(\eta)$, spatial components for the constituent particles' momenta, \emph{i.e.}, by taking 
\begin{equation}
\label{Pieta}
P_i^m=E_0(U^m+\eta V_i^m)
\, ,
\quad
\text{with} 
\quad
V_i \cdot  U=0
\, ,
\quad
\sum_i V_i^m=0
\, .
\end{equation}
Sparing the reader the uninspiring details, we find that the term in the right hand side of~(\ref{deltaXi3}) also contributes to this higher order,  but two more terms show up, proportional to 
\begin{equation}
\label{twomoreterms}
\sum_i R_{\triangle}(P_i,P) \Xi_i^m
\, ,
\quad
\sum_i E^{-3} (P_i \cdot P) R_\triangle(P_1+P_2,P_i)\Xi_i^m
\end{equation}
respectively (with $R_\triangle(A,B) \equiv R_{abcd}X^aY^bA^cB^d$)--- their particular contribution to this order may be computed by substituting~(\ref{Pieta}) in the above equation. 

It is worth pointing out that, in most astrophysical situations, the $\calO(\epsilon^3 \eta)$ displacements in (\ref{deltaXiav}), (\ref{twomoreterms}), are quite small, compared to the typical size of the objects involved. However, the situation may be dramatically reversed if we contemplate extending an analysis similar to this even to the regime 2. Moreover, there are some fundamental issues that emerge from the discussion of this section, one of which is discussed next. 

\subsubsection{Implications for the renormalization program}

We should emphasize the fundamental role that associativity plays in many of the standard theoretical techniques in physics. For instance, the renormalization program within the Wilsonian approach \cite{Wilson}, is based on the idea of integrating up the degrees of freedom that are associated with very high energies, or very short scales, to obtain an effective characterization that is relevant at lower energies. Evidently, for this process to have any hope of leading to trustable results, it is essential that when integrating from the high energy scale $\Lambda^{(1)}$ directly to the lower energy scale $\Lambda^{(3)}$, we obtain  the same result as that one would reach by first integrating from $\Lambda^{(1)}$ to the intermediate scale $\Lambda^{(2)}$, and then integrating the resulting effective theory from $\Lambda^{(2)}$ to the low energy scale $\Lambda^{(3)}$. In fact such path independence lies at the core of the renormalization group techniques which are an essential aspect of the whole effective theory approach. It should be clear that, when associativity fails, all bets are off regarding such path independence.

To see this more clearly imagine we have a spacetime with metric $g_{ab}$, and a set of $6N$ particles with masses $m_{i}$ and world lines $\gamma_{i}$, where the subscript $i$ labels the particle, and we proceed to integrate out the short scale by dividing the collection into $2N$ groups of $3$ particles. This leads to $2N$ world lines representing the CoM of effective particles, on a spacetime with effective metric $ \tilde{g}_{ab}$, that have masses $\tilde{m}_{j}$ and that move along their CoM world lines $\tilde{\gamma}_{j}$, where now $j=1,\ldots,2N$. Observe that we can associate to these situations an effective energy-momentum tensor using, for example, equation (\ref{EMT4}). Moreover, we can reiterate the process dividing the $2N$ effective particles into $N$ sets of particle pairs and integrating out another scale of the system. The point is that one could instead start by dividing the collection in $N$ sets of 6 particles each, integrating out and representing the situation by $N$ composite particles and the two procedures lead to two inequivalent energy-momentum tensors which, however, are supposed to describe the same physical situation at the same coarse graining level.
Note that this issue, as described here, concerns the scale transition of classical systems, however, something similar can be expected to occur when dealing with such change of scale in quantum mechanical regimes such as quantum field theory.

The issue that needs to be considered is the following: suppose we start with a characterization of the matter degrees of freedom at the quantum gravity scale $\Lambda^{\rm Planck}$, and assume we integrate out the fluctuations associated with very high energy scales. In that way we recover a characterization of the situation at hand in terms of an effective spacetime metric and a quantum field theory on that curved spacetime, with an effective energy-momentum tensor and with characteristic cutoff scale $\Lambda^{\rm GUT}$. On the other hand, we would like to recover the effective description at, say, the electroweak scale $\Lambda^{\rm EW}$. The question is whether we would obtain the same result by integrating directly from $\Lambda^{\rm Planck}$ to the low energy scale $\Lambda^{\rm EW}$. We do not, at this point, carry out any detailed calculation but rather argue on the basis of what has been found that the answer is likely going to be negative. Also, we want to point out that similar concerns have been raised in references \cite{matarrese1,matarrese2} within the cosmological context.

The lack of associativity of the CoM that we have found, and which shows that, in general, the effective geodesics one obtains depend on the manner one integrates out the fundamental constituents of our test objects, suggests that the resulting metrics that would be obtained in the two ways, are generically different, and that those differences depend strongly on the characteristics of the fluctuations, or more specifically, on the behavior of the fundamental degrees of freedom at the $\Lambda^{\rm Planck}$ scale.
 
One of the fundamental suitability tests for a quantum gravity theory is the recovery, in an appropriate limit, of GR. Our findings above indicate that perhaps the test itself needs to be substantially reformulated.
Of course, the problem could be bypassed if one could define an observer-independent CoM world line which is associative, but at the moment we see no path in this direction. To recover associativity one would have to sacrifice observer independence and take a relational point of view, as it is done in reference \cite{OurPaper}. In this case one uses the simultaneity hypersurface associated with some particular observer, regardless of the object under study. However, such an observer dependent CoM would not be a suitable starting point for recovering an observer-independent metric characterization of spacetime.

\subsection{Emergent canonical structures}

There are well known arguments that combine quantum mechanics and GR which conclude that there must be fundamental localizability limits associated with black hole formation once sufficient energy is confined in a small enough region \cite{BlackHoleFormation}. In this regard we would like to put forward a simple argument suggesting that the issue ought to be revised. Suppose we manage to concentrate a given amount of energy in a region that, as seen in frame $S$, is close to the black hole forming regime. Now, for an observer boosted with respect to $S$, with boost factor $\gamma$, the size of the region where the energy is confined is shrank by  $\gamma$ while the energy concentrated in that region increases by the same factor. Thus, if $\gamma$ is large enough, the black hole formation bound is reached, for the boosted observer, and the two observers disagree in their expectations of black hole formation. This, in our view, illustrates the need to investigate in more detail issues that, like this one, lie at the interface between gravity and quantum theory, with observer independent methods.

On the other hand, as it is well known, the position operator plays a very important role in quantum mechanics, and there is no satisfactory definition of this operator in quantum field theories. Now, since the CoM we have been considering seems to be applicable to situations involving the matter fields energy-momentum tensor, it is tempting to generalize the recipe to quantum field theories, where the energy-momentum tensor is an operator, and to try to define a position operator following that CoM construction. We observe that, in generic situations, this proposal requires dealing with expectation values of the energy-momentum tensor, therefore, determining the CoM world line faces several additional complications as compared to the situation involving the energy-momentum tensor of a classical $N$-particle system \cite{EMTensor1,EMTensor2}. 

In this subsection we consider the manner in which the discussions in the previous sections impinge on quantum aspects of the matter description. Our focus is on the fact that the quantization process is designed to reproduce the symplectic structure with the replacement of Poisson brackets by quantum mechanical commutators. Thus, modifications of the Poisson brackets would have repercussions in the construction of the matter quantum theory. We now argue that the Poisson structure should be modified if we use the variables characterizing the CoM of the extended objects.

In reference \cite{Pryce} it was pointed out that, in special relativity, the observer-independent CoM has nonvanishing Poisson brackets among its different components. In fact, these brackets depend on the total momentum and spin of the extended object. This nontrivial expression stems from the fact that the relativistic CoM weighs the particles' position by their energies rather than their masses, which, appropriately allows massless particles to contribute to the location of the CoM. As the particles' energies depend on their momenta, the CoM can be thought of as a function of the position and momenta of the constitutive particles of the extended object. Moreover, the resulting form of the Poisson brackets is remarkably similar to expressions found in rather different approaches \cite{chryss}. It is also noteworthy that, in contrast to most works on noncommutative geometry, this noncommutativity arises naturally and it is fully compatible with Lorentz invariance. The point is that the noncommutativity here is tied to the objects used in probing spacetime rather than to the underlying spacetime itself.

Coming back to the Poisson brackets, at first sight it seems straightforward to generalize the calculation of reference \cite{Pryce} to GR for an extended object comprised by $N$ free point-like particles by using the corresponding Poisson brackets of the point-like particles. However, to actually perform such a calculation is technically daunting and we can only sketch some qualitative conclusions. We can separate the issues related to this calculation into two classes: fundamental and technical. The fundamental difficulties stem from the fact that the Poisson brackets are related, by definition, with phase space functions. In turn, phase space is associated with a particular spacetime foliation (i.e., a notion of time) that allows one to describe the evolution of such phase space functions. In contrast, the construction of CoM in curved spacetime involves quantities living in different leaves of a given foliation. However, if the evolution of each particle conforming the extended object is known, we can translate this information to phase space variables to calculate Poisson brackets. 

Moreover, it is unclear what quantities can be taken to be the canonical pairs. In this regard, it seems natural to use the foliation associated with the rest frame itself, put the origin of the coordinates at the actual CoM, and use a normal Riemann frame associated to that point. Then, the spatial components of the vectors whose exponential maps intersect the particles' world lines and the spatial components of the momenta transported to this point are good candidates to be canonical pairs. 

Regarding the technical difficulties, note that when calculating Poisson brackets one needs to consider changes of the state of the system since these brackets involve taking derivatives with respect to the canonical variables. However, working with the CoM recipe we are using, these changes, in turn, imply changes in the hypersurface on which the parallel transport is done. Furthermore, one has to be careful to ensure one is working with truly independent variables, namely, if one's coordinate frame is tied to the CoM, then, when considering a variation in the state of the system corresponding to, say, one of the constitutive particles, the location of the CoM is modified, thus changing the vectors pointing to the other particles. Therefore, to actually perform the calculation one would need to use a particular fiducial foliation and a fixed coordinate system where all the degrees of freedom are independent. A choice would be to use the foliation associated with the CoM rest frame and to place the origin at the CoM world line, but allowing it to change (i.e., to be displaced from the origin) when variations are considered. In this sense it seems interesting to perform a perturbative analysis similar to that presented in section \ref{nonassoc} where only the first corrections due to curvature are sought --- this is still technically challenging since one has to make sure that the approximations are properly implemented especially after taking derivatives, and it is thus left as the subject of a future work.  We hope that it is still possible to use symmetry arguments and argue \textit{a priori} what is the dominant curvature correction.

Nevertheless, even before calculating explicitly the Poisson brackets we are interested in, we can say that, as a correction to Pryce's  result, there will be  terms proportional to the spacetime curvature. This, we expect, is due to the fact that, as we described above, when taking variations of the particles' positions and momenta, both the CoM and the rest-frame hypersurface are changed, and, when comparing with the state before the variation, one needs to bring the objects from those surfaces back onto the original foliation, resulting in a holonomy due to the parallel transport involved. The outcome is that the curvature of the underlying spacetime will modify the Poisson structure with respect to that of the classical theory in nontrivial ways. Then some interesting questions arise. For instance, what is the full Poisson bracket algebra satisfied by the effective degrees of freedom,  and what type of action would give rise to such symplectic structures?

\section{Conclusions}
\label{Diss} 

We have considered some obstacles that will be faced when trying to recover classical geometry from a fundamental theory of quantum gravity. We argued that, when recovering the classical spacetime notions of GR, one would have to rely on the identification of matter states that could play the role of free particles and, moreover, that the effective geometry should be read off from the covariant CoM world lines characterizing such objects. This has lead us to examine the characteristics of the CoM of extended objects, which was done in a simpler setting where the background spacetime is assumed as given. 

We pointed out the relevance for such a program of the well-known fact that the CoM world lines do not follow the geodesics of the background spacetime. This, together with the requirement  that we should have a scheme where the equivalence principle is built in at the fundamental level into the very notion of spacetime geometry, suggests that, in extracting the effective spacetime geometry, one should identify \textit{a priori} the CoM world line of a free extended object as a geodesic of an effective geometry. However, as we have shown, even this relatively simple step in the recovery of an effective geometry is filled with complications. In particular, we noted that there are difficulties in finding all the components of the effective connection because several identical extended objects are needed to probe the given region in all spacetime directions. In addition, we have also described a method for obtaining an effective curvature tensor by considering the relative acceleration between the CoM world lines of two nearby extended objects, and found that the results depend strongly on the detailed characteristics of such objects.

On the other hand, the fact that the dominant part of many of the effects we considered depends on the total spin of the extended object, as already noted by Papapetrou, suggests that one should use extended objects with zero total spin to characterize the effective metric geodesics.  However, when the dominant term is insensitive to such a choice, higher multipolar momenta of the extended body would come into play in an essential way. One may then chose to use as probes, at a given scale, objects with vanishing, up to a certain order,  higher multipolar momenta. This suggests that, at least in the context of the averaging problem and the quest to obtain effective metrics at a given scale, one could use the following approach: given an underlying spacetime metric, an averaging scale, and an accuracy, consider extended objects with proper spatial extent in that scale, and determine what are the minimal number of energy-momentum tensor multipolar momenta that need to vanish to ensure that the center of mass does not depend on the probes within the corresponding subclass of test extended bodies. One may then use such world lines to define the averaged spacetime metric at the corresponding scale. In practice, this would be a reasonable recipe to address the averaging problem in cosmology and astrophysics.

Moreover, the difficulties in extracting a spacetime geometry with extended objects suggest that the task would be even harder when dealing with quantum probes. In fact, when considering the possible application of the prescription involving extended bodies with vanishing multipolar momenta, as described above, several problems appear right away. First and foremost, the fact that we need to start with an underlying spacetime metric to define the probe's multipolar momenta, makes it evident that this prescription is unsuitable to connect the regimes 1 and 2. And even in other regimes, we have to deal with the fact that, given the lack of associativity of the construction, it is impossible to have a recipe that works regardless of the initial scale. This suggests that one needs to rely on something like a smallest possible scale. However, since it is hard to point out to a fundamental length scale without coming into conflict with Lorentz Invariance \cite{Collins:2004,Dowker:2004,Mattingly:2005}, this approach seems unsuitable at the fundamental level. In other words, the proposal seems inadequate in the quest to recover the GR spacetime from a quantum gravity theory. This raises the possibility that, in a regime where the quantum properties of the particles cannot be neglected, the geometrical language itself might cease to have meaning, since it may be impossible to properly define it.

We have also shown that the covariant definition of the CoM is not associative in the sense that it cannot be used to represent the parts into which one divides an extended object. That is, in order to calculate the CoM of the complete extended object, all the individual particles have to be identified as such, as neglecting their internal compositeness would lead to incompatible results. This can have practical consequences when using the CoM to characterize an extended object in regions with high curvature, but it is also interesting at the conceptual level because it suggests that the CoM is only well defined if fundamental particles exist. This nonassociativity of the CoM construction also has important consequences at the quantum level. In particular, it would cast serious doubts on the whole renormalization program for a quantum field theory that can be considered as effective and emergent from the underlying theory of quantum gravity since the notion that one can integrate over certain degrees of freedom and then consider the effective variables obtained from such integration as new fundamental variables, is at the basis of the renormalization group approach.

We also considered briefly the issue of how to generalize the calculation of the Poisson brackets of the components of the center of mass to curved spacetimes and how this indicates a problem when recovering the quantum characterization of the matter degrees of freedom. Interestingly, this problem seems to persist in situations where one believes that the underlying quantum nature of spacetime can be neglected. The expected nontrivial appearance of the spacetime curvature in the commutators of the free particles implies that there could be curvature effects on the quantum theory of ``point particles" extracted from the underlying theory of quantum gravity. This indicates, in turn, that at the quantum level one can expect obstacles when recovering the GR prescription of minimal coupling. Effects characterizing a granular spacetime structure could then show up \cite{Phenomenology1,Phenomenology2,Phenomenology3,Phenomenology4}. Also, since the fundamental matter degrees of freedom cannot be taken as infinitely localized, it is unclear how to operationally state the equivalence principle.
 
There is still a lot of work to be done along the lines described in this manuscript. In particular, it is intriguing to see if it is possible to generalize the construction of the center of mass to situations where matter is described in the realm of quantum field theory in curved spacetimes. It seems that this would be a required first step in order to define a covariant position operator which could be used to determine some canonical world lines that could be used to define an effective geometry. Naturally, such a program would encounter similar difficulties to those we have described in this work. In fact, it can be expected that those difficulties would appear aggravated simply because, in that context, one needs to regularize and renormalize the energy-momentum tensor. In view of the nonassociativity we have discussed above, those steps indicate that we should question the meaning of an effective geometry which would be tied to the motion of real physical matter probes. On the other hand, it is doubtful whether something that is not intimately related with the behavior of such matter probes should be rightfully described as the physical spacetime geometry.

After facing all these problems, we are tempted to conclude that the underlying quantum gravity theory should be such that, when extracting classical GR, nothing like what we presented here arises. In fact, it seems that such an underlying theory has to be of a rather different nature from the usual approaches and, in particular, it should not involve any fundamental spacetime discreetness. 
We do not want, at this point, to argue that the approach we have described in this paper represents the only possible path, or that the obstacles we have found are truly insurmountable. On the other hand, we don't see an alternative approach that both, bypasses the difficulties we have encountered, and still provides a satisfactory recovery of GR from quantum gravity.

\section*{Acknowledgements}
We acknowledge P. Aguilar for insightful comments and T. Koslowski and the late J. D. Bekenstein for pointing out some interesting references. 

\bibliography{CM} 
\bibliographystyle{ieeetr}

\end{document}